# Beyond prior knowledge: The predictive role of knowledge-building in Tutor Learning

Tasmia Shahriar [0000-0003-0199-7757], Mia Ameen [0009-0008-8541-2480], Aditi Mallavarapu 3[0000-0002-2906-1478], Shiyan Jiang[0000-0003-4781-846X], and Noboru Matsuda3[0000-0003-2344-1485]

North Carolina State University, Raleigh NC 27695, USA
(tshahri,fameen,amallav,sjiang24,noboru.matsuda)@ncsu.edu

**Abstract.** When adopting the role of a teacher in learning-by-teaching environments, students often struggle to engage in knowledge-building activities, such as providing explanations and addressing misconceptions. Instead, they frequently default to knowledge-telling behaviors, where they simply dictate what they already know or what to do without deeper reflection, thereby limiting learning. Teachable agents, particularly those capable of posing persistent follow-up questions, have been shown to encourage students (tutors) to shift from knowledge-telling to knowledge-building and enhance tutor learning. Tutor learning encompasses two interrelated types of knowledge: conceptual and procedural knowledge. Research has established a bidirectional relationship between these knowledge types, where improvements in one reinforce the other. This study investigates the role of knowledge-building in mediating the bidirectional relationship between procedural and conceptual learning. Our findings revealed a stable bidirectional relationship between procedural and conceptual knowledge, with higher post-test scores observed among students who engaged in knowledge-building, regardless of their procedural and conceptual pre-test performance. This suggests that knowledge-building serves as a crucial mechanism bridging the gap between students with low prior knowledge and higher conceptual and procedural learning gain.

**Keywords:** Learning-by-teaching, Large Language Models, Constructive Tutee Inquiry, Knowledge-building

## 1 Introduction

Engaging in knowledge-building activities are recognized as an effective way to promote students learning while solving problems [1-4]. *Knowledge-building* activities include reflecting on own understanding by revisiting concepts, providing explanations to make sense of solution steps, and recovering from misconceptions or knowledge gaps [5, 6]. Research shows that students often struggle to engage in knowledge-building when adopting the role of a teacher in the learning-by-teaching environment [3, 7-9]. In such cases, students frequently adopt a *knowledge-telling* approach with their teachable agents, where they primarily rephrase known information without analyzing or synthesizing it, thereby limiting their learning. Teachable agents in the learning-by-teaching environment, particularly those capable of posing persistent follow-up questions, can encourage students, who we refer as *tutors*; to shift from a knowledge-telling mode to a knowledge-building mode and facilitate tutor learning [2, 10, 11].

Tutor learning can be examined through three fundamental types of knowledge: *Conceptual*, *Procedural* and *Procedural Flexibility* [12-14]. *Conceptual knowledge* is defined as the knowledge of abstract principles and the interrelations between concepts, while *procedural knowledge* is defined as the knowledge of procedures. *Procedural flexibility,* on the other hand, is defined as the knowledge of solving a problem in multiple ways and determining which method is most efficient [15, 16]. Procedural knowledge facilitates the effective execution of techniques, while conceptual knowledge enables the transfer of learning across different contexts and fosters deeper comprehension [17]. The interaction between conceptual and procedural knowledge forms a well-established bidirectional relationship, where gains in one type of knowledge can reinforce and enhance the other. This interplay is essential in the development of procedural flexibility skills [18, 19].

Despite the recognized importance of these three types of knowledge, the extent to which knowledge-building in a learning-by-teaching environment influences their development remains underexplored. Knowledge-building in a learning-by-teaching environment has been shown to strengthen procedural knowledge [10, 11]. However, does engaging in knowledge-building within such settings also foster the development of conceptual knowledge and procedural flexibility? If so, how does this influence the bidirectional relationship between procedural and conceptual knowledge?

The goal of the current paper is to investigate the role of knowledge-building activities on tutors' learning conceptual, procedural and procedural flexibility within the learning-by-teaching environment. Our main contributions are: (I) We replicated the bidirectional relationship between procedural and conceptual knowledge in a learning-by-teaching environment;  and (II) We found that engaging in knowledge-building was positively correlated with increased gains in conceptual knowledge, procedural knowledge, and procedural flexibility, and the effect remains significant after controlling tutors' prior knowledge.

## 2    Related Work

Teachable agents have been explored as a way to enhance tutor learning by encouraging explanation and reflection. However, existing systems vary in how effectively they promote knowledge-building.

Curiosity Notebook is a learning-by-teaching platform where tutors learn to classify objects by teaching a conversational teachable agent [20]. The TA relies on a template-based questioning framework that does not adapt to tutors' knowledge gaps, limiting opportunities for cognitive conflict and knowledge-building [21]. Additionally, the effectiveness of its template-based questioning framework in fostering knowledge-building remains unexamined in prior research.

A more advanced system, TeachYou, integrates a Large Language Model (LLM)-based teachable agent, Algobo, which teaches programming concepts like binary search by shifting between receiving help and questioning the tutor. It persistently asks tailored

"why" and "how" follow-ups to elicit knowledge-building responses [22]. While effective, the study focused primarily on procedural knowledge and did not examine its impact on conceptual learning or the broader relationship between the two within learning-by-teaching environment. Additionally, its evaluation relied on qualitative analysis and surveys rather than pre-posttest comparisons. This limits our ability to assess the system's impact on tutor learning.

Beyond teachable agent's questioning systems, prior work has established the bidirectional relationship between conceptual and procedural knowledge in mathematical learning; with improvements in procedural knowledge often supporting improvements in conceptual knowledge as well as vice versa [18, 23, 24]. However, the extent to which knowledge-building mediate this relationship in a learning-by-teaching environment remains an open question.

In sum, there is a gap in the current literature regarding how questioning frameworks can simultaneously foster both procedural and conceptual learning through knowledge-building, in the context of a learning-by-teaching environment. Our study addresses this gap in the literature by conducting extended data analysis of tutors' interactions with ExpectAdapt, a novel teachable agent questioning framework, to reveal the dynamics between knowledge-building and procedural/conceptual learning.

## 3   APLUS: The Learning-By-Teaching Environment

In our study, we used APLUS (**A**rtificial **P**eer **L**earning **U**sing **S**imStudent) as our learning-by-teaching environment where tutors (i.e., students who are learning by teaching) assist SimStudent (the teachable agent) how to solve linear algebraic equations [25]. **Figure 1** displays the user interface of APLUS. Whenever tutor enters a linear equation to teach (**Figure 1-a**), SimStudent tries to solve one step at a time by consulting its knowledge base that consists of production rules once learned like, "if [*conditions*] hold then perform [*a solution step*]." In APLUS, the *solution step* allows four basic math operations: *add*, *subtract*, *multiply*, and *divide by* a term. If SimStudent has a production that can apply, it seeks feedback from the tutor. If the tutor agrees, it proceeds to the next step. If the tutor disagrees, it asks a focal question, "Why am I wrong?". The tutor is expected to provide their textual explanation in a chat box (**Figure 1-c**). If SimStudent does not have a production to apply, it requests the tutor to demonstrate the next step. After tutor demonstrates the solution step, it asks another focal question, "Why should we do it?". After tutors respond to the focal questions, SimStudent asks further follow-up questions based on a questioning framework called *ExpectAdapt* [11], which is discussed in the next section.

Apart from teaching, tutors can quiz SimStudent to evaluate how well SimStudent has learned thus far by observing the SimStudent's performance on the quiz. Quiz topics include one-step equations (level 1), two-step equations (level 2), equations with variables on both sides (level 3), and a final challenge that contains equations with variables on both sides (level 4) (**Figure 1-g**). SimStudent works on a single quiz level at a time. Upon successfully passing a level, the subsequent level is unlocked.

Tutors may also review the resource tabs that include problem bank, unit overview, introduction video and worked out examples at any time (**Figure 1-d**). The teacher agent, Mr. Williams (**Figure 1-h**), provides on-demand, voluntary hints on how to teach. For example, if the tutor repeatedly teaches one-step equations, Mr. Williams might provide the hint, "SimStudent failed on the two-step equation. Teaching similar equations will help him pass that quiz item."

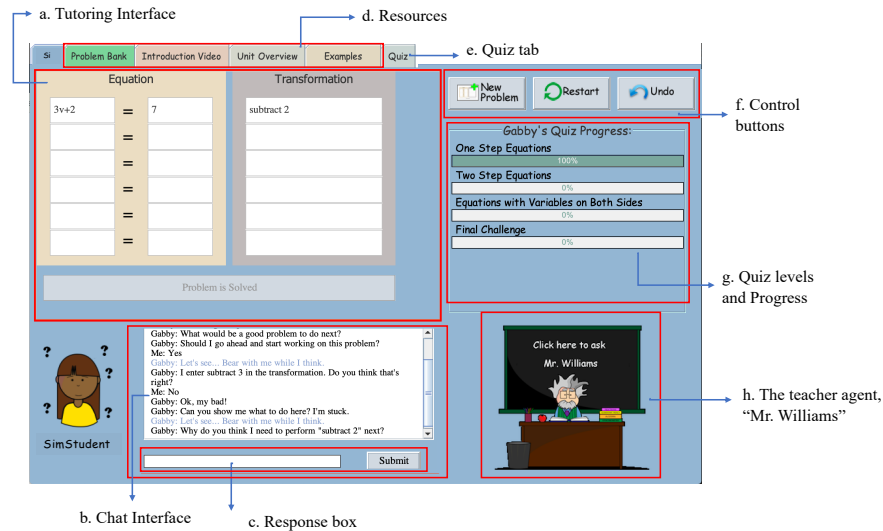

**Figure 1:** APLUS interface with SimStudent shown in the bottom left corner.

## 4 Intelligent Teachable Agent's Questioning Framework

To effectively elicit knowledge-building in tutors, we integrated the Expectation-tailored Adaptive questioning framework, or ExpectAdapt, a follow-up questioning framework into APLUS [11]. The following section outlines the motivation behind its development and the details of its implementation.

### 4.1 Motivation

In a learning-by-teaching environment, tutors often exhibit a tendency to neglect or inadequately respond to a teachable agent's questions; Graesser et al. [13] introduced a persistent follow-up questioning technique to improve the quality of students' answer in a classroom setting. Subsequent research has reinforced the finding, demonstrating that tutors who provide elaborated responses using conceptual terms learn significantly more than tutors who cannot, irrespective of their prior knowledge [2, 10]. Existing literature also argues that tutors switch from providing knowledge-telling to knowledge-building responses by hitting impasses or moments when they realize they do not know something or need to double-check their understanding [3, 9, 26]. Such moments are highly improbable to attain without proper scaffolding on the tutors' responses. These findings underscore the need for teachable agents that actively generate follow-up questions to prompt knowledge-building.

Research further emphasizes that effective questioning must be tailored to individual tutors' understanding, while simultaneously pushing their cognitive boundaries, maintaining discourse coherence, context relevance, and inextricably bound to the conceptual content of the subject matter [27, 28]. Prior studies using scripted follow-up questions, which did not adapt to tutors' responses, facilitated procedural but not conceptual learning. These findings underscore the need for an adaptive approach that considers conversational context when generating follow-ups.

ExpectAdapt addresses these gaps by generating adaptive follow-up questions that guide tutors toward elaborated response [11]. It generates follow-up questions tapping on the concepts of the elaborated response that tutor has not conveyed yet. The framework is closely aligned with AutoTutor's expectation & misconception-tailored dialogue (aka, *EMT* dialogue) [29]. Unlike AutoTutor, which relies on scripted authoring tools requiring extensive human expertise, ExpectAdapt leverages the in-context learning capabilities of Large Language Models to detect conceptual gaps within a tutor's response. This process mirrors how a student (TA in this context) clarifies ambiguities in a textbook with a teacher's guidance avoiding corrective questions.

Prior research on ExpectAdapt demonstrated its effectiveness in engaging tutors in knowledge-building. Tutors who answered its follow-up questions learned more while teaching fewer problems, simply by spending more time on responding to the questions [11]. Given this success, we integrate ExpectAdapt into APLUS for this study. The next section outlines how ExpectAdapt works.

### 4.2 ExpectAdapt

ExpectAdapt consists of three stacked Large Language Models (LLMs): (1) Expected Response Generator, (2) Alignment Detector, and (3) Follow-up question generator.

The system operates by generating follow-up questions after a tutor responds to a focal question. First, the Expected Response Generator internally generates an expected response reflective of knowledge-building relevant to the focal question. The Alignment Detector then evaluates the tutor's response by comparing it to the expected response and categorizes it into one of three classifications: (1) aligned, (2) not aligned, or (3) unable to detect. If the response is aligned, the Follow-up Question Generator formulates a targeted question that addresses any missing or underdeveloped components of the expected response while maintaining the conversational context. This mimics a teachable agent's effort to clarify ambiguities and bridge knowledge gaps. However, if tutors' response is not aligned or the alignment detector is unable to determine alignment, the system halts follow-up questioning to prevent overly corrective or irrelevant prompts. ExpectAdapt generates a maximum of three follow-up questions per focal question to ensure sustained engagement without overwhelming the tutor.

All three LLM modules use OpenAI API for the GPT-3.5-turbo model. To optimize question generation, ExpectAdapt leverages various prompt engineering techniques, including few-shot demonstrations using chain-of-thought [18-21], role prompting [22], and adding extra context in the prompt [23, 24]. Additionally, the term "teachable agent" is deliberately avoided in the prompts, replaced by "student.". This choice is grounded in the hypothesis that LLMs may encounter difficulties in assuming the role of a teachable agent, a scenario presumably less prevalent in their pretraining datasets—LLMs excel with more common terms from pretraining [25]. Further implementation details and functional analysis of ExpectAdapt can be found in [11].

## 5 Methods

We next describe the study designed to examine the impact of knowledge-building on procedural and conceptual knowledge and their bidirectional relationship, as well as the tutors' learning gain as a result of the intervention.

### 5.1 Participants

A total of 23 middle school students (6th to 8th grade) from various regions across the United States participated in our study. Participation was voluntary, and informed consent was obtained from both the students and their parents. Participants were compensated $15 per hour for their involvement in the study.

### 5.2 Study Design

The study employed a pretest–intervention–posttest design. During the intervention, participants interacted with SimStudent and responded to tailored follow-up questions generated by the ExpectAdapt framework. The study was conducted in a hybrid format, allowing participants to choose between in-person attendance or remote participation via Zoom. For remote participants, the application was accessed using lab computers through Zoom screen sharing and remote-control access to ensure a consistent study environment.

Participants engaged in the study for one hour per day over five consecutive days, following a structured schedule: On Day 1, they completed a 30-minute pretest to establish prior knowledge. Day 2 began with an introductory video explaining the purpose and functionality of the application, after which participants started interacting with the application. Days 3 and 4 involved continued engagement with the application. Finally, on Day 5, participants completed a 30-minute isomorphic post-test.

### 5.3 Test Materials

We adopted the test administered by Rittle-Johnson, Star [28] to assess equation-solving skills in middle school students. The pre and post-test consist of three modules: (1) Procedural Skill Test (PST) (2) Conceptual Knowledge Test (CKT) and (3) Procedural Flexibility Test (PFT). In our study we included the items reported in Schneider, Rittle-Johnson [18] that covered one-step equation, two-step equation and equations with variable on both sides. The highest score any participant could achieve in the overall test, PST, CKT and PFT, are 18, 14, and 2, respectively. Details of the assessment items are shown in **Table 1**. We utilized a binary scoring system for each test items, i.e., answers were marked strictly as either correct or incorrect. Test scores are normalized as the ratio of participant's score to the maximum score.

Table 1: Sample Items for Assessing Procedural, Conceptual and Flexibility Knowledge

| Problem Type | Sample items |
| --- | --- |
| *Procedural Skill Test* (PST) ($n = 18$) | |
|    Equation-solving ($n = 10$) | Solve for $y$; $16 + 5y = -8 - y$ |
|    Effective next step ($n = 8$) | Evaluate if the step would simplify the equation; <br> Equation: $5a - 3 = 2$ <br> Step: Dividing both sides by 3 <br> Options: Correct / Incorrect / I do not know |
| *Conceptual Knowledge Test* (CKT) ($n = 14$) | |
|    Equal sign ($n = 2$) | Without solving the equation assess if the equation $39x - 4 = 10$ has the same solution as $10 = 39x - 4$? <br> Options: Yes / No / I do not know |
|    Like terms ($n = 5$) | $-j$ and $7j$ like terms with each other. <br> Options: Correct / Incorrect / I do not know |
|    Coefficient ($n = 5$) | 4 is a coefficient of the term $-4x$. <br> Options: Correct / Incorrect / I do not know |
|    Negative term ($n = 2$) | Does $-4 + x$ has the same quantity as adding $x$ to $-4$? <br> Options: Yes / No / I do not know |
| *Procedural Flexibility Test* (PFT) ($n = 2$) | Solve this equation in two different ways: $2x + 8 = 4x + 6$ |

### 5.4 Analysis

The following components supported our analyses of the study results:

**KBR Classifier.** To understand if tutors commit knowledge-building or knowledge-telling while interacting with SimStudent, we used a response classifier proposed in [10] to automatically classify tutors response as either indicative of knowledge-building or knowledge-telling. Responses that are tagged as indicative of knowledge-building are called *knowledge-building responses* (KBR) and responses that are tagged as indicative of knowledge-telling are called *knowledge-telling responses* (KTR). The classifier proposed in [10] tags responses as KBR or KTR based on four key dimensions: sentence formation (ill-formed vs. well-formed), relevancy (relevant vs. irrelevant), information content (why-informative vs. what/how-informative), and intonation (descriptive vs. reparative). Well-formed, relevant, why-informative, and descriptive or reparative responses containing domain-specific key terms are defined as KBR and responses lacking any of the above categories are defined as KTR.

In our analysis, we used the percentage of KBR (**%KBR**) as the ratio of tutor responses classified as KBR to the total number of responses (KBR+KTR) generated by tutors during their interaction with SimStudent.

**Path Analysis Model.** Path analysis is always theory-driven; the same data can describe many different causal patterns. Thus, it is essential to have an a priori idea of the causal relationships among the variables under consideration. As our baseline model, we adopt the model presented by Schneider, Rittle-Johnson [18] to test the bidirectional relationship between procedural and conceptual knowledge and their combined effect on procedural flexibility. The model includes direct effect of $PST_{pre}$ on $PST_{post}$, $CKT_{post}$ and $PFT_{post}$; as well as direct effects of $CKT_{pre}$ on $CKT_{post}$, $PST_{post}$, and $PFT_{post}$. This model aims to test how procedural and conceptual prior knowledge influence their respective post-test outcomes and procedural flexibility post-test outcomes. We further include direct effect of $PST_{pre}$ and $CKT_{pre}$ on *%KBR* and direct effect of *%KBR* on $CKT_{post}$, $PST_{post}$, and $PFT_{post}$. Our specified path analysis model is shown in **Figure 2**. The model displayed adequate fit to the data; $\chi^2(1) = 0.58$, $p = .45$, the Comparative Fit Index (CFI) is 1.0, Tucker-Lewis Index (TLI) is 1.1, the Root Mean Square Error of Approximation (RMSEA) is 0.0 and the Standardized Root Mean Squared Residual (SRMR) is 0.02.

**Qualitative Analysis.** Due to the small sample size (only 20 data points), extensive comparative quantitative analysis (e.g., a regression analysis or a two-way ANOVA) was not feasible. Instead, we conducted a qualitative analysis of tutor responses to explore how some low prior knowledge students learned to engage in knowledge-building. We focused on the responses of low-prior, high-%KBR students to examine the potential impact of follow-up questions in eliciting knowledge-building activities.

## 6 Results

### 6.1 Tutors have a significantly higher conceptual and procedural test score at post-test

**Table 2** summarizes tutors' pre and post test scores on conceptual, procedural and procedural flexibility test. The results of paired t-tests comparing pre- and post-test performance for each category indicate that there is a significant improvement in conceptual and procedural knowledge following the intervention. However, for the procedural

**Table 2: Performance Summary: %Correct ($M \pm SD$)**

|  | Conceptual *CKT* | Procedural *PST* | Procedural Flexibility *PFT* |
|---|---|---|---|
| Pre-test | $0.78 \pm 0.18$[a] | $0.71 \pm 0.32$[b] | $0.54 \pm 0.50$[c] |
| Post-test | $0.90 \pm 0.13$[a] | $0.83 \pm 0.26$[b] | $0.63 \pm 0.46$[c] |

a: $t(22) = -4.47$, $p < 0.001$; b: $t(22) = -2.71$, $p < 0.05$; c: $t(22) = -1.45$, $p = 0.16$.

flexibility test, the improvement from pre to post-test was not found to be statistically significant.

*These results suggest that the intervention was particularly effective in enhancing conceptual and procedural knowledge, while procedural flexibility may require further targeted support.*

### 6.2 Conceptual and procedural knowledge have stable bidirectional relations

From the regression coefficient of our path analysis model, shown in **Figure 2**, conceptual knowledge at pre-test is a predictor of procedural skill at post-test ($\beta = .34, p < .05$). Similarly, procedural skill at pre-test is a predictor of conceptual knowledge at post-test ($\beta = .16, p < .05$). The regression coefficient in our findings are in the same range as the findings reported by Rittle-Johnson, Siegler [30] and Schneider, Rittle-Johnson [18].

Schneider, Rittle-Johnson [18] also reported that conceptual and procedural knowledge at pre-test contributed significantly to procedural flexibility at post-test. While our data support procedural knowledge at pretest as a predictor of procedural flexibility ($\beta = .56, p < .05$), our current data did not reveal that conceptual knowledge at pretest as a predictor of procedural flexibility.

*These results suggest a stable bidirectional predictive relationship between procedural and conceptual knowledge that the prior competency of one type of knowledge has non-trivial impact on learning another type of knowledge through learning by teaching with knowledge-building dialogue.*

### 6.3 Percentage of knowledge-building response is a strong predictor of conceptual, procedural and procedural flexibility score

Our path analysis model revealed two interesting findings: (1) The regression coefficient from conceptual and procedural pre-test score to %KBR was not significant ($CKT^{Pre}: \beta = .33, p = .24$; $PST^{Pre}: \beta = .18, p = .25$). This suggests that *tutors' conceptual and procedural prior knowledge do not significantly contribute to %KBR.* (2) %KBR is a predictor of conceptual, procedural and procedural flexibility score at post-test.

*These findings suggests that tutors, regardless of their prior knowledge, who had a higher percentage of knowledge-building response to the teachable agent's questions ended up with higher conceptual, procedural and procedural flexibility score in the post test. This indicates that the ExpectAdapt framework effectively facilitated knowledge-building, which in turn enhanced conceptual/procedural learning.*

### 6.4 Qualitative Analysis

**Figure 3** illustrates a representative example of how tutor-tutee dialog may help students learn to produce KBR more efficiently over time. In **Figure 3-a**, we observe the student's initial attempts at responding to SimStudent's inquiries about a specific problem. Despite having low priors, the tutor improves the quality of their responses with each follow-up, culminating in a KBR by the second follow-up. In a later interaction with the teachable agent **(Figure 3-b)**, the student produces a KBR after just one follow-up, demonstrating their growing ability to engage in knowledge-building.

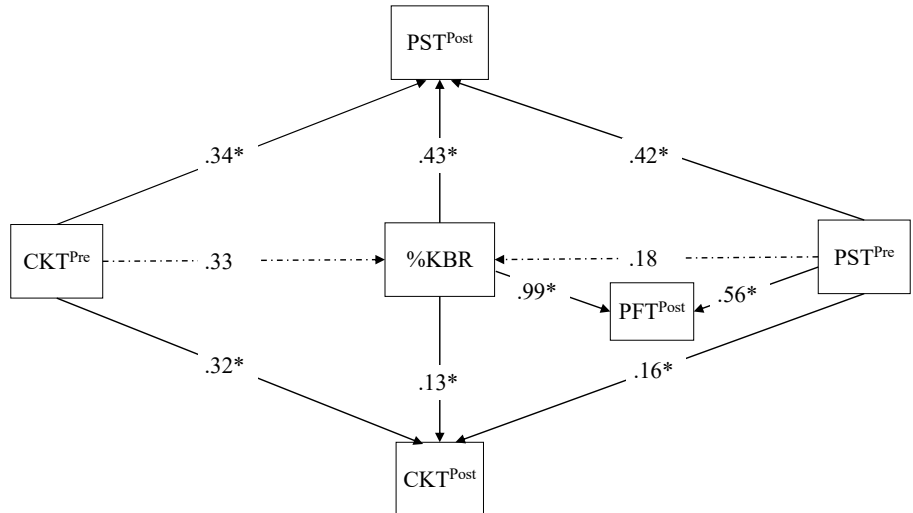

**Figure 2:** Regression path coefficients of the described model to measure the relation between percentage of KBR (%KBR) and conceptual, procedural and procedural flexibility scores. All estimated coefficients significant with $p < .05$ are shown in * superscript. Non-significant links were omitted, except for the two dashed link from procedural and conceptual prior; $PST^{pre}$ and $CKT^{pre}$ to %KBR for reference.

## 7 Discussion

Our data revealed that tutors learned both conceptual and procedural knowledge from the intervention; however, no such effect was observed for procedural flexibility. One possible explanation is the limited assessment scope, as we only used two items to measure this knowledge type. Future studies could refine the assessment by incorporating a broader and more diverse set of items to capture procedural flexibility more effectively. Additionally, it is possible that procedural flexibility requires more extensive scaffolding or repeated exposure to varied problem-solving scenarios, which were not sufficiently emphasized in this intervention.

We replicated the bidirectional relationship between procedural and conceptual knowledge, as demonstrated by Schneider, Rittle-Johnson [18], within a learning-by-teaching environment. Path analysis revealed that procedural pre-test score predicts tutors' conceptual post-test score, while conceptual pre-test score predicts procedural post-test score. While these findings suggest predictive relationships, they do not establish causality. Future research should consider controlling for tutors' procedural and conceptual knowledge during experimental design to better understand the underlying mechanisms driving the bidirectional relationship. In contrast to Schneider, Rittle-Johnson [18] report that both procedural and conceptual knowledge predict procedural flexibility, our findings only supported procedural knowledge as a predictor of procedural flexibility. This discrepancy may be attributed to limitations in the assessment items used to measure procedural flexibility.

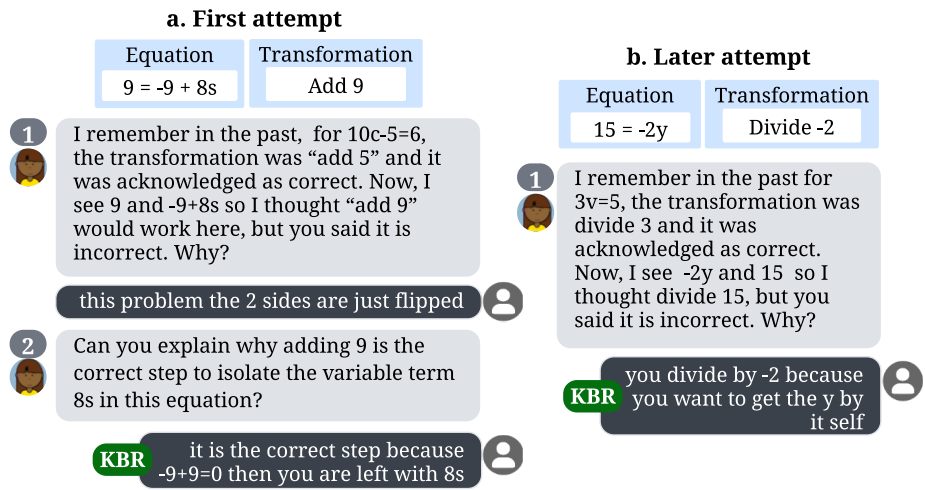

**Figure 3:** An example of tutor-tutee dialog between SimStudent (left) and a low-prior-knowledge tutor (right). The current problem and proposed solution step are displayed on the top of the diagram. Follow-up questions are numbered sequentially. Student's knowledge-building responses are labeled as KBR.

Additionally, our path analysis revealed that the percentage of knowledge-building responses (%KBR) significantly predicts conceptual, procedural, and procedural flexibility post-test scores. A key concern was whether this effect was primarily driven by students with high prior knowledge who were naturally more capable of generating knowledge-building responses. This concern aligns with existing research in the learning-by-teaching paradigm since prior knowledge plays a crucial role in tutor learning. Rodrigo, Ong [31] reported that tutors with insufficient foundational knowledge may struggle to benefit from tutoring, whereas those with adequate prior knowledge are more likely to experience meaningful learning gains.

However, our path analysis model showed that prior knowledge—both procedural and conceptual—was not a significant predictor of %KBR. This suggests that active engagement in knowledge-building is a stronger determinant of test score than prior knowledge itself. Notably, tutors across all levels of prior knowledge who exhibited a higher percentage of knowledge-building responses in response to the teachable agent's questions demonstrated greater improvements in conceptual, procedural, and procedural flexibility scores on the post-test.

This finding is particularly compelling because it challenges the assumption that only students with high prior knowledge can benefit from learning-by-teaching. Instead, it highlights the critical role of engagement in knowledge-building as a mechanism for learning. Even tutors with lower prior knowledge can achieve greater test scores if they actively construct and articulate their understanding. Our qualitative analysis revealed instances of dialogue between SimStudent and the low prior tutor, where low prior tutors learned to engage in knowledge-building due to repeated follow-up questions. These results underscore the importance of designing tutoring environments that encourage knowledge-building interactions, ensuring that all learners—regardless of prior knowledge—can maximize their conceptual and procedural development.

## 8   Conclusion

We found that students learned both conceptual and procedural knowledge during the intervention where they responded to teachable agent's follow-up questions. We also found a stable bidirectional relationship between conceptual and procedural knowledge. Specifically, students with higher procedural knowledge prior to the intervention tended to achieve higher conceptual test scores after the intervention, while those with higher conceptual knowledge at the outset showed greater improvements in procedural test scores. This bidirectional relationship emphasizes the interconnected nature of these two forms of knowledge, suggesting that improving one can reinforce and enhance the other.

Additionally, our analysis revealed that engaging in knowledge-building activities was positively correlated with higher scores in procedural, conceptual, and procedural flexibility after the intervention. What stands out in these findings is the observation that students with lower prior knowledge were also able to generate a higher percentage of knowledge-building responses, as prior test scores did not predict the percentage of knowledge-building response generated by students. This challenges the assumption that only students with high prior knowledge benefits from learning-by-teaching, highlighting the role of follow-up questions in fostering knowledge-building, even for students with low prior knowledge.

While these findings are promising, it is important to acknowledge that the relationships we observed are correlational, not causal. Our path analysis shows associations within our data, but it does not establish causality. To better understand these effects, we plan to conduct a randomized controlled trial in future research.

**Acknowledgment:** This research was supported by the Institute of Education Sciences, U.S. Department of Education, through grant No. R305A180517 and National Science Foundation grant Nos. 2016966 and 2112635 to North Carolina State University.  The opinions expressed are those of the authors and do not represent views of the Institute or the U.S. Department of Education and NSF.